\documentclass[procedia]{easychair}
\pdfoutput=1

\newcommand{\ang}{\textup{\r{A}}}
\newcommand{\MAE}{\textup{MAE}}

\title{Ab-Initio Study on the Hard Magnetic Properties of MnBi}
\titlerunning{Ab-Initio Study on MnBi}

\author{
    Peter Toson\inst{1}
    \and
    Ahmad Asali\inst{1}
    \and
    Gregor A. Zickler\inst{1}
    \and
    Josef Fidler\inst{1} \\
}

\institute{
    Institute of Solid State Physics, Vienna University of Technology, Vienna, Austria\\
    \email{peter.toson@tuwien.ac.at} \\
}

\authorrunning{Toson, Asali, Zickler and Fidler}

\begin{document}

\maketitle

\keywords{MnBi, Permanent Magnets, First Principles}

\begin{abstract}
We have studied the hard magnetic properties of the low-temperature phase of MnBi with first principle calculations based on the density functional theory. The calculations have been carried out on two distinct unit cell configurations \emph{MnBi} and \emph{BiMn} with the element in the unit cell origin named first. Our results show that these configurations are not equivalent and that \emph{MnBi} describes the system better near $T = 0K$ and the \emph{BiMn} configuration describes the system better for $T > 300K$. The magnetic moments of both configurations agree well with experimental measurements considering both spin and orbital contributions. At high temperatures the magneto-crystalline anisotropy energy increases with increasing unit cell volume and reaches a maximum of $2.3 MJ/m^3$ and a $c/a$ ratio of $1.375$.

%\textbf{Ref: ID 2539, Paper Number: 2372.00, Paper Reference: TU.H-P110}
\end{abstract}

\section{Introduction}

The low-temperature phase (LTP) of MnBi has interesting properties suitable for permanent magnetic applications \cite{rama_rao_anisotropic_2013}. At low temperatures its spin configuration is in-plane but at $T = 80K$ a spin reorientation takes place and the magneto-crystalline anisotropy energy (MAE) increases with increasing temperature up to $2.1kJ/m^3$ at $T = 500K$. At $T = 593K$ a first-order phase transition occurs reducing the $c/a$ ratio \cite{wijn_1.1.3.2.4_1988}. These properties motivate a systematic study of the magnetic properties of MnBi in dependence on the change of lattice parameters with temperature.

\begin{figure}[htb]
	\centering
		\includegraphics[width=0.90\textwidth]{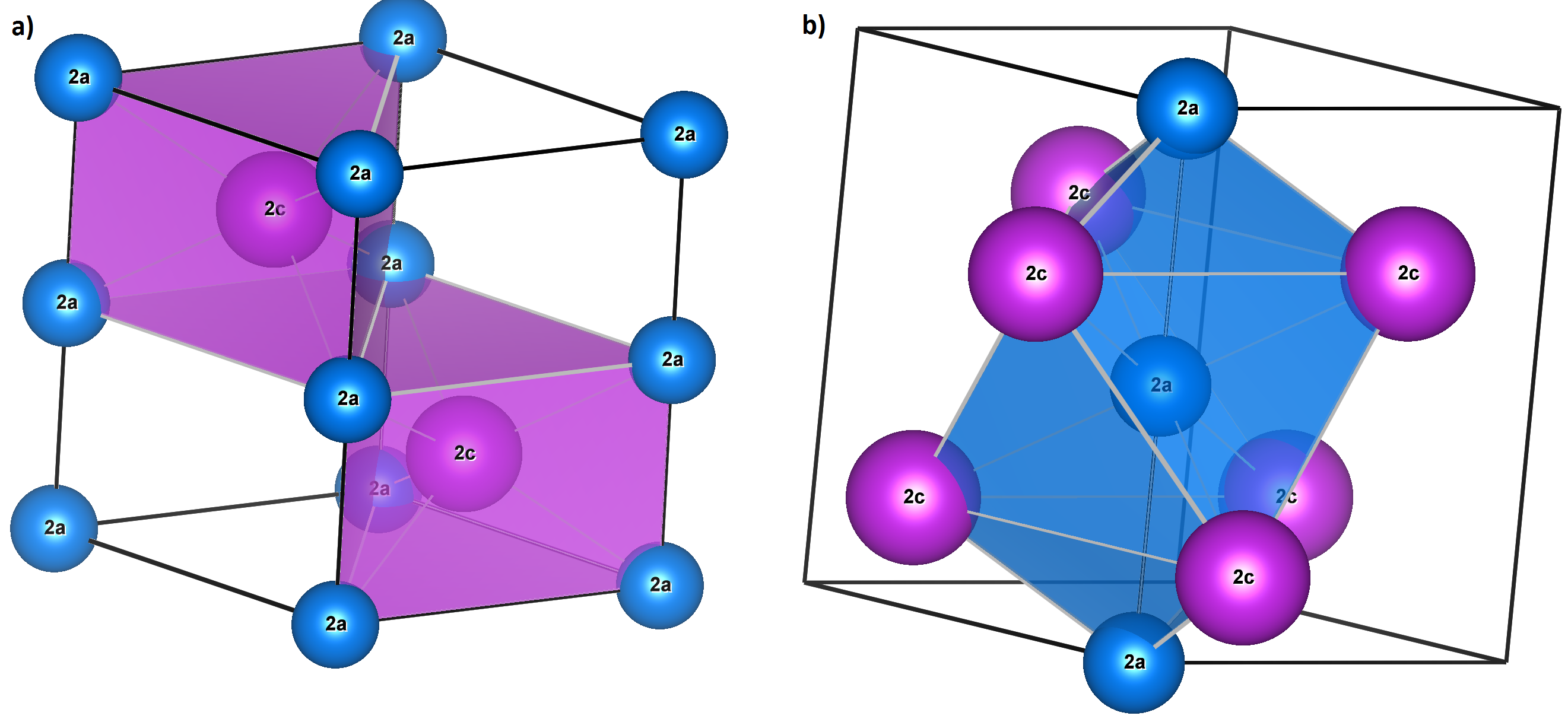}
	\caption{Atomic environments in the MnBi unit cell. (a) Trigonal prism around $2c$ cite. (b) Distorted square anti-prism surrounding $2a$ site. Figures created with VESTA \cite{momma_vesta_2011}.}
	\label{fig:fig1-atomicenvoronments}
\end{figure}

The LTP phase of MnBi is a hexagonally close packed structure with \textbf{ABAC} stacking and described by space group 194 $P6_3/mmc$. The \textbf{A} layers have the Wyckoff position $2a$ the alternating \textbf{B} and \textbf{C} layers $2c$. The $2a$ and $2c$ positions have different atomic environments, which means that exchanging Mn and Bi positions leads to different systems (see fig. \ref{fig:fig1-atomicenvoronments}). For clarity, the configuration with Mn at the $2a$ site will be called \emph{MnBi} and the configuration with Bi at the $2a$ site will be called \emph{BiMn} throughout the paper. Sources for both configurations can be found in literature (e.g. BiMn: \cite{chen_contribution_1974, coey_magnetism_2013}, MnBi: \cite{gobel_properties_1976, shanavas_theoretical_2014})

\section{Method}

Ab-initio calculations based on the density functional theory (DFT) have been performed. The central statement of the DFT is that the total energy of a many-electron system can be expressed as an functional of the electron density. This leads to the formulation of the Kohn-Sham equation, a Schr\"odinger-like equation on a continuous electron density \cite{kohn_self-consistent_1965}. The Kohn-Sham equation can be solved in a self-consistent manner.

The calculations have been carried out with the WIEN2k package, a full-potential, augmented plane wave code \cite{schwarz_solid_2003, blaha_wien2k_2001}. As exchange-correlation potential the generalized gradient approximation in the formulation of Perdew, Burke and Ernzerhof (PBE-GGA) has been used \cite{perdew_generalized_1996}. The main advantage of GGA over the local (spin) density approximation L(S)DA is that not only the electron density, but also its first derivative is taken into account.

Spin-polarized, scalar-relativistic calculations on both MnBi configurations have been performed until convergence with a criterion of $10^{-9}Ry$ on a $25 \times 25 \times 15$ k-mesh has been reached. After that, the relativistic effects such as the orbital contraction of Bi s-states and the spin-orbit coupling (SOC) have been introduced. The MAE is the difference of the total energies with SOC in easy and hard direction: $\MAE = E_{[100]} - E_{[001]}$. It is worth noting that the results depend on the SOC initialization method and the resulting crystal symmetries. While the error due to incorrect symmetries is small compared to the total energy of the system ($10^4 Ry$), it is in the order of typical $\MAE$ values ($10^{-5} Ry$). Orbital contributions to the total magnetic moments have been determined afterwards by calculating the density matrix.

\section{Results}

\subsection{Lattice Optimization}

Based on the lattice parameters reported in \cite{wijn_1.1.3.2.4_1988} calculations to find the equilibrium lattice parameters of \emph{MnBi} have been performed. Optimizing unit cell volume and $c/a$ ratio yields $a = 4.3408 \ang$ and $c = 5.7142 \ang$ with $c/a = 1.3164$. Both lattice parameters deviate from the experimentally measured values ($a = 4.285 \ang$, $c = 6.113 \ang$ in \cite{chen_contribution_1974}) but are in good agreement with other DFT calculations \cite{shanavas_theoretical_2014, ming-qiu_ab_2000}. 

The equilibrium lattice parameters correspond to $T = 0K$. The MAE of the \emph{MnBi} system was determined to be $-0.11 MJ/m^3$ which is in better agreement with experimental measurements ($\MAE = -0.2 MJ/m^3$ at $T = 4K$, \cite{rama_rao_anisotropic_2013, wijn_1.1.3.2.4_1988}) compared to other DFT calculations ($\MAE = -2.1 MJ/m^3$, \cite{shanavas_theoretical_2014, ravindran_magnetic_1999}). Calculations on the \emph{BiMn} system wrongfully predict an uniaxial anisotropy with $\MAE = +1.55 MJ/m^3$ at $T = 0K$.

\subsection{Magnetic Moments}

The total spin magnetic moment per \emph{MnBi} unit cell is $7.37 \mu_B$. The main contribution to the spin moment come from the Mn atoms ($3.77 \mu_B$) especially from d electrons ($3.68 \mu_B$). The spins of the Bi atom are aligned anti-parallel to the Mn spins lowering the total spin moment. The interstitial moments ($0.07 \mu_B$) are contributions from electrons outside the muffin-tin radius and cannot be assigned to a single atom. Detailed per-orbital contributions are summarized in table \ref{tab:SpinMomentsMnBi}.

These results are in good agreement with other DFT calculations, but lower than experimental results with $\mu_{Mn} =  3.82 - 4.25 \mu_B$ \cite{rama_rao_anisotropic_2013, shanavas_theoretical_2014}. Considering the orbital moments as well (see table \ref{tab:OrbMomentsMnBi}) this experimental value can be reached without the introduction of an phenomenological Hubbard $U$ potential. The magnetic moments per atom can be increased by 5\% by variation of the lattice parameters, however, the magnetic moment per volume stays constant.

Magnetic moments of the \emph{BiMn} configuration are between 9\% and 12\% higher than \emph{MnBi} systems throughout the examined range of lattice parameters and agree with the upper bound of experimental values (see tables \ref{tab:SpinMomentsBiMn} and \ref{tab:OrbMomentsBiMn}).

\begin{table}[htbp]
	\centering
		\begin{tabular}{lrrrrr}
            \hline
            & \multicolumn{1}{c}{\textbf{s}} & \multicolumn{1}{c}{\textbf{p}} & \multicolumn{1}{c}{\textbf{d}} & \multicolumn{1}{c}{\textbf{f}} & \multicolumn{1}{c}{\textbf{sum}} \\
            \hline
                \textbf{Mn}  & 0.05770	&  0.02721	& 3.68389	& -0.00008	&  3.76872 \\
                \textbf{Bi}	 & 0.02100	& -0.15420	& 0.01055	&  0.00432	& -0.11833 \\
                \textbf{interstitial} & &           &           &           &  0.07154 \\
                \textbf{sum} & 0.07870	& -0.12699	& 3.69444	&  0.00424	&  3.72193 \\
            \hline
		\end{tabular}
	\caption{Spin Moments of \emph{MnBi} at $T = 0K$ in $\mu_B$}
	\label{tab:SpinMomentsMnBi}
\end{table}

\begin{table}[htbp]
	\centering
		\begin{tabular}{lrrrr}
            \hline
            & \multicolumn{1}{c}{\textbf{p}} & \multicolumn{1}{c}{\textbf{d}} & \multicolumn{1}{c}{\textbf{f}} & \multicolumn{1}{c}{\textbf{sum}} \\
            \hline
                \textbf{Bi}	 & -0.00011 &  0.07802 &  0.00020 &  0.07811 \\
                \textbf{Mn}	 & -0.00413 & -0.00015 & -0.00010 & -0.00438 \\
            \hline
                \textbf{sum} & -0.00424 &  0.07787 &  0.00010 &  0.07373 \\
            \hline
		\end{tabular}
	\caption{Orbital Moments of \emph{MnBi} at $T = 0K$ in $\mu_B$}
	\label{tab:OrbMomentsMnBi}
\end{table}

\begin{table}[htbp]
	\centering
		\begin{tabular}{lrrrrr}
            \hline
            & \multicolumn{1}{c}{\textbf{s}} & \multicolumn{1}{c}{\textbf{p}} & \multicolumn{1}{c}{\textbf{d}} & \multicolumn{1}{c}{\textbf{f}} & \multicolumn{1}{c}{\textbf{sum}} \\
            \hline
                \textbf{Bi}  & 0.02437 & -0.07656 & 0.01122 & 0.00558 & -0.03539 \\
                \textbf{Mn}  & 0.08713 &  0.04057 & 3.81380 & 0.00024 &  3.94174 \\
                \textbf{interstitial}  &  &       &         &         &  0.28596\\
            \hline
                \textbf{sum} & 0.11150 & -0.03599 & 3.82502 & 0.00582 &  4.19231\\
            \hline
		\end{tabular}
	\caption{Spin Moments of \emph{BiMn} at $T = 0K$ in $\mu_B$}
	\label{tab:SpinMomentsBiMn}
\end{table}

\begin{table}[htbp]
	\centering
		\begin{tabular}{lrrrr}
            \hline
            & \multicolumn{1}{c}{\textbf{p}} & \multicolumn{1}{c}{\textbf{d}} & \multicolumn{1}{c}{\textbf{f}} & \multicolumn{1}{c}{\textbf{sum}} \\
            \hline
                \textbf{Bi}  & -0.00754 & -0.00046 & -0.00052 & -0.00852 \\
                \textbf{Mn}  & -0.00171 &  0.10225 &  0.00026 &  0.10080 \\
            \hline
                \textbf{sum} & -0.00925 &  0.10179 & -0.00026 &  0.09228 \\
            \hline
		\end{tabular}
	\caption{Orbital Moments of \emph{BiMn} at $T = 0K$ in $\mu_B$}
	\label{tab:OrbMomentsBiMn}
\end{table}

\subsection{Temperature Dependence of MAE}

The temperature dependence of the magnetic properties have been calculated by variation of the lattice parameters. Only thermal expansion has been considered in these calculations, other thermal effects like temperature smearing of electron bands or phonons have been ignored. Measurements show that in the temperature range from $300 K$ to $700 K$ the $c/a$ ratio varies between $1.35$ and $1.45$ and the unit cell volume increases from $97 \ang ^3$ to $101 \ang ^3$.

\begin{figure}[htbp]
	\centering
		\includegraphics[width=0.60\textwidth]{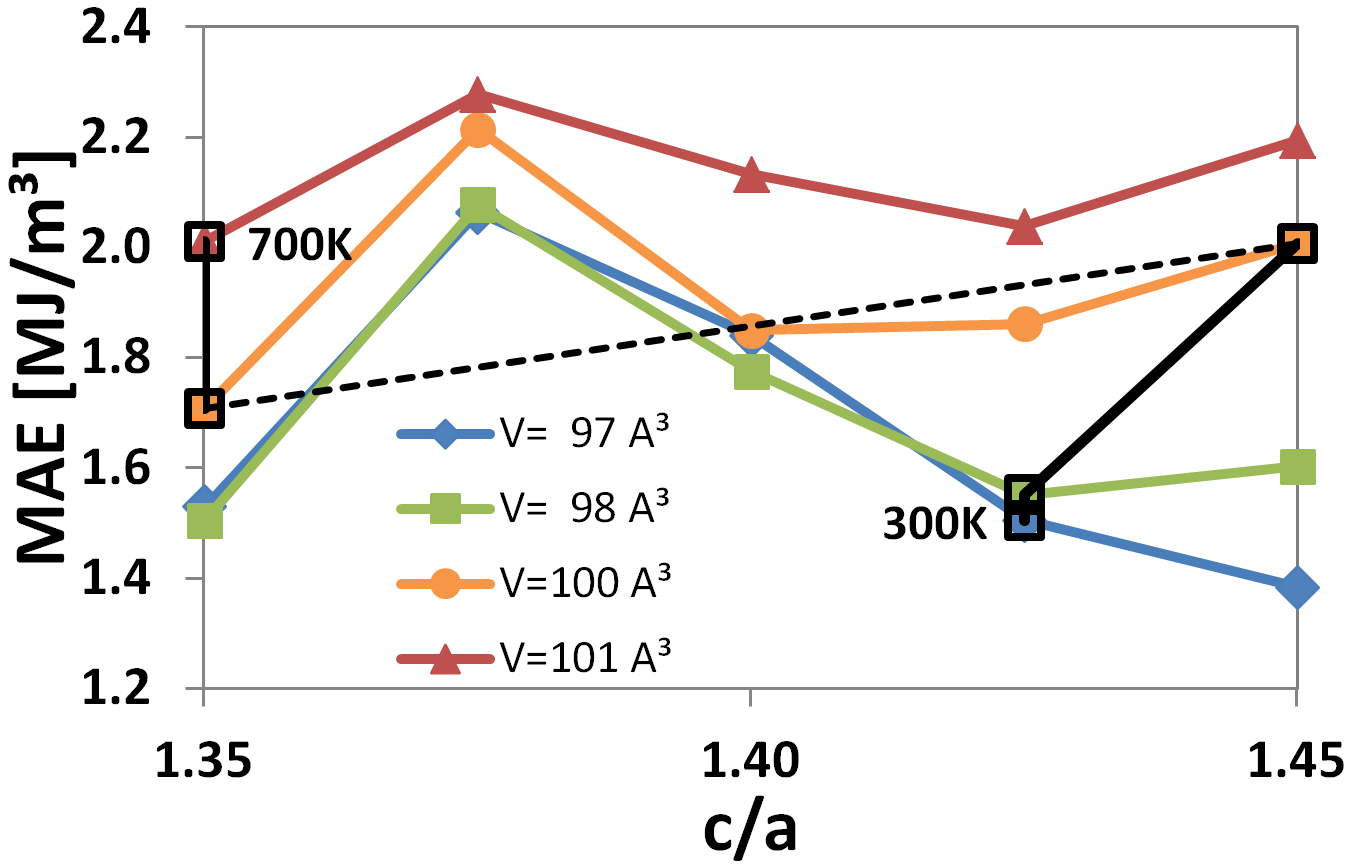}
	\caption{MAE dependence on unit cell volume and $c/a$ ratio, \emph{BiMn} configuration. The black line shows the temperature dependence of volume, $c/a$ ratio and $\MAE$. In the dotted region the first order transition takes place leading to $c/a$ reduction.}
	\label{fig:fig2-variation}
\end{figure}

The change of lattice parameters is not sufficient to change the sign of MAE of the \emph{MnBi} system, but the values obtained from the \emph{BiMn} system are in good agreement with experimental results (see fig. \ref{fig:fig2-variation}). With increasing temperature (unit cell volume) the MAE of the \emph{BiMn} system increases. The MAE is maximized with $c/a$ ratios around $1.375$ and the influence of the $c/a$ ratio is higher at lower volumes.

\section{Conclusion}

We have successfully calculated the magnetic properties of LTP MnBi with two non-equivalent lattice configurations: \emph{MnBi} (Mn at the origin) describes the properties around $0 K$ and \emph{BiMn} (Bi at the origin) describes the high-temperature properties from $300 K$ to $700 K$. Both configurations have been confirmed experimentally in literature. The results suggest that thermal expansion alone cannot explain the unusual temperature-dependent magnetic properties of MnBi, but they could be a consequence of a transition between \emph{MnBi} and \emph{BiMn} configuration. The high-temperature MAE increases with increasing unit cell volume and is maximized by $c/a$ ratios of $1.375$.

\section*{Acknowledgements}
The financial support of the EC-FP7 projects REFREEPERMAG (280670) and ROMEO (309729) is acknowledged. The computational results presented have been achieved by using the Vienna Scientific Cluster VSC-2.

\label{sect:bib}
\bibliographystyle{unsrt}
\bibliography{bib}

%\textbf{DONT FORGET THE UMLAUTE!!! G\"obel, 1974}

\end{document}